\documentstyle[aps,12pt]{revtex}

\input{epsf}

\begin{document}

\title{Thermal and Dynamic Effects in Langevin Simulation of
       Hysteresis in Nanoscale Pillars}
\author{Gregory Brown$^{a}$, 
        M.A. Novotny$^{a}$, 
        and P.A. Rikvold$^{a,b}$}
\address{
$^{a}$
School of Computational Science and Information Technology,
Florida State University, Tallahassee, Florida 32306-4120\\
$^{b}$
Center for Materials Research and Technology, and Department of
Physics,
Florida State University, Tallahassee, Florida 32306-4350
}

\date{\today}
\maketitle

\begin{abstract}
Dynamic quantities related to hysteresis have been measured in
micromagnetic simulations of single-domain nanoscale magnets at
nonzero temperature.  The hysteresis-loop area and magnetization-field
correlation display the characteristics of resonance, and the
resonance frequency is found to be temperature dependent. The
period-averaged magnetization displays symmetry breaking at high
frequencies.
\end{abstract}


The common theme in hysteretic systems is a nonlinear, irreversible
response that lags behind an applied force. While different physical
mechanisms may cause this behavior, in many interesting situations it
results from a system with two local free-energy minima separated by a
free-energy maximum. For instance, the crystalline or shape anisotropy
of a magnetic particle may create a barrier that significantly
interferes with the reversal mechanism whereby the magnetization
changes from the unfavorable alignment antiparallel to the applied
field to the equilibrium alignment parallel to the
field. Historically, hysteresis experiments have been conducted under
quasistatic conditions that minimize thermal and dynamic effects. With
new technology making applications of magnetization reversal of
nanoscale magnets on the nanosecond time scale important, a thorough
understanding of these effects is needed. Here we present results from
dynamic micromagnetic modeling of hysteresis in nanoscale pillars at
nonzero temperatures.

Our numerical model is based on iron nanopillars fabricated using
STM-assisted chemical vapor deposition~\cite{Wirth:98,Wirth:99}. 
Following the one-dimensional model for magnetic pillars discussed by
Boerner and Bertram~\cite{Boerner:97}, the nanoscale magnets are
decomposed into a one-dimensional stack of $17$ cubes with side
$7.2\,{\rm nm}$. Each cube has a magnetization density ${\bf M}({\bf
r}_i,t)$ and a local magnetic field ${\bf H}({\bf r}_i,t)$, with ${\bf
r}_i$ the center of the $i$-th cube. The damped precessional motion of
each magnetization vector is given by the Landau-Lifshitz-Gilbert
(LLG) equation~\cite{Brown:63,Aharoni}
\begin{equation}
\label{eq:llg}
\frac{ {\rm d} {\bf {M}}({\bf r}_i) }
     { {\rm d} {t} }
 =
   \frac{ \gamma_0 }
        { 1+\alpha^2 }
   {\bf {M}}({\bf {r}}_i)
 \times
 \left (
   {\bf {H}}({\bf {r}}_i)
  -\frac{\alpha}{M_s} {\bf {M}}({\bf {r}}_i) \times 
                      {\bf {H}}({\bf {r}}_i)
 \right )
\;,
\end{equation}
where the electron gyromagnetic ratio is $\gamma_0$$=$$1.76$$\times$$10^7
{\rm Hz/Oe}$~\cite{Aharoni}, and $\alpha$$=$$0.1$ is a phenomenological
damping parameter. The numerical details have been given
elsewhere~\cite{Brown:MMM00,Brown:MMM01,Brown:Opus}. Parameters
corresponding to bulk iron were used: saturation magnetization density
$M_s = 1700 \, {\rm emu/cm}^3$ and exchange length $l_{\rm e} = 3.6
\, {\rm nm}$. The local magnetic field ${\bf H}({\bf r}_i,t)$ is a
linear superposition of terms corresponding to exchange, magnetostatic
interactions, the applied field, and a stochastic field representing
thermal fluctuations. The latter term is governed by the
fluctuation-dissipation theorem, for which we use the form derived
for isolated magnetic particles~\cite{Brown:63,Brown:Opus}. The
stochastic integration~\cite{Brown:Opus} of Eq.~(\ref{eq:llg}) was
performed using a first-order Euler scheme with $\Delta
t$$=$$5$$\times$$10^{-14} \, {\rm s}$.

Thermal fluctuations are essential in this model because the
free-energy barrier can be surmounted using thermal energy momentarily
``borrowed'' from the surroundings. For the spatially nonuniform
reversal mechanisms considered
here~\cite{Boerner:97,Brown:MMM00,Brown:Opus}, this barrier crossing
occurs at the ends of the pillar. The observed hysteretic behavior
depends on the interplay of three time scales: the mean nucleation
time, the time required for the subsequent growth of the nucleated
volumes to fill the particle, and the period of the applied field.

To simulate hysteresis loops, an external field 
${\bf H} = \hat{\bf z} H_0  \cos{(2\pi\nu t)}$ 
is applied, with $H_0$$=$$2000\,{\rm Oe}$. The results are presented
for the reduced quantities $m(t)$, the $z$-component of the
magnetization for the entire magnet, and $h(t)$, the applied magnetic
field, both normalized to have maximum values of unity.

The correlation between $m$ and $h$ is measured by 
\begin{equation}
\label{eq:B}
B = \frac{\pi\nu}{2} \oint m(t) h(t) {\rm d}t
\;,
\end{equation}
where the integral is taken over one hysteresis loop. The results are
presented vs. $\nu$ in Fig.~1 for $T$$=$$20\,{\rm K}$ and $100\,{\rm
K}.$ The bars shown for $T$$=$$100\,{\rm K}$ are {\em not error bars},
instead they represent the standard deviation of the probability
density for $B$. An example of $m(t)$ is shown as the solid curve in
the inset of Fig.~1, along with $h(t)/10$ which appears as the dotted
curve. In the high-frequency limit the magnetization usually does not
switch, and $B$ is near zero. At low frequencies the magnetization
switches nearly every period, and $B$ approaches unity. The quantity
$B$ corresponds to the reactive part of the nonlinear response
function~\cite{Sides:98}, and the lowest-frequency zero crossing of
$B$ in Fig.~1 is associated with resonance.

The dissipative part of the nonlinear response is the hysteresis-loop
area,
\begin{equation}
\label{eq:A}
A = -\frac{1}{4} \oint m(h) {\rm d}h
\;.
\end{equation}
The measured average values of $A$ are shown vs.\ $\nu$ in Fig.~2 for
both temperatures. At high frequencies the magnetization is unable to
respond to the applied field, and the hysteresis loop area is nearly
zero as most of the time $m(t)$ oscillates with a small amplitude
about one or the other of the degenerate values of the zero-field
magnetization. In the low-frequency (or quasi-static) limit the
magnetization usually switches at small values of $h(t)$, and as a
result $A$ is again near zero. At intermediate frequencies $A$
displays a maximum near the resonance frequency.  The maximum in $A$
and the low-frequency zero crossing in $B$ agree to about $15\%$. The
temperature dependence of the results is mostly due to the different
average nucleation times associated with the different
temperatures. Similar simulations in two-dimensional Ising magnets
have shown that the phenomena seen here are associated with stochastic
resonance~\cite{Sides:98}.

Another interesting quantity to consider is the period-averaged
magnetization,
\begin{equation}
\label{eq:Q}
Q = \nu \oint m(t) {\rm d}t
\;,
\end{equation}
which corresponds to a dynamic order parameter for the system.
Histograms are shown for several applied-field frequencies at
$T$$=$$100\,{\rm K}$ in Fig.~3. At high frequencies the histogram has
two maxima, corresponding to the magnetization being oriented
predominantly in the positive or negative direction during the entire
hysteresis loop. At low frequencies the histogram is centered around
zero since the magnetization switches during every half-period of the
applied field. Previously, a phase transition in this dynamic order
parameter has been seen in Ising models of nanoscale
magnets~\cite{Sides:PRL,Sides:99,Korniss:01,Fuji:01} when many nucleation
events contribute to each reversal of the magnetization. In these
models, this phase transition occurs near the frequency where $B(\nu)$
has its high-frequency zero crossing. However, the detailed analysis
and large statistical sampling required to make such a determination
has not been performed for the present simulations.

In summary, micromagnetic simulations of nanoscale magnets at nonzero
temperatures and gigahertz frequencies display the characteristics of
temperature-dependent resonance. Future work exploring the
connections to stochastic resonance and dynamic phase transitions is
planned.

This work was partially supported by the Ames Laboratory, which is
operated for the U.S. Department of Energy by Iowa State University
under Contract No. W-7405-82, by the Laboratory Directed Research and
Development Program of Oak Ridge National Laboratory, managed by
UT-Battelle, LLC for the U.S. Department of Energy under Contract
No. DE-AC05-00OR22725, and by the U.S. National Science Foundation
through grant DMR-9871455. Computer resources were provided by the
Department of Physics at Florida State University.

\newpage

\begin{figure}[tb]
\center{\epsfxsize 4 in \epsfbox{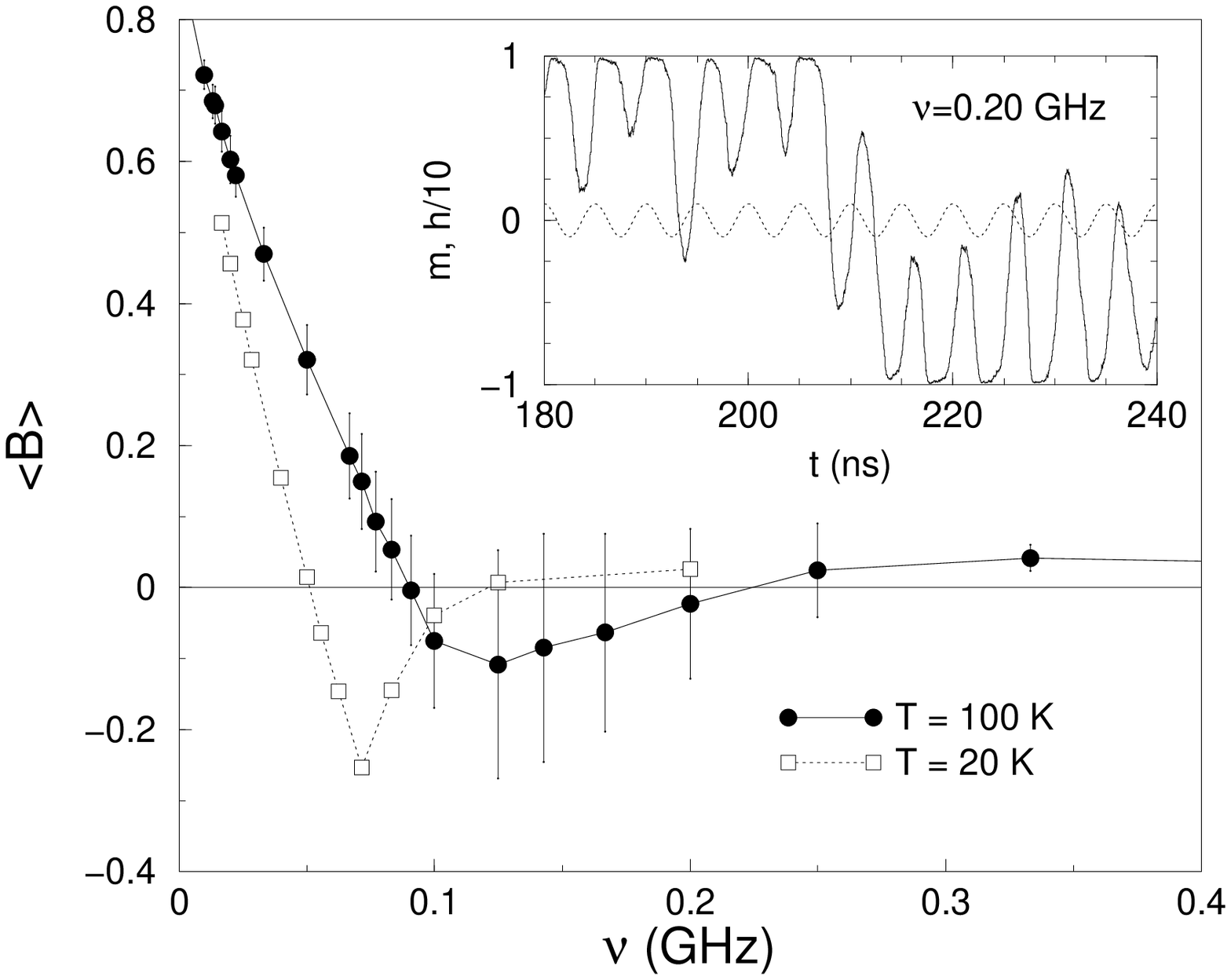}}
\caption[]{The correlation $B$, defined in Eq.~(\protect{\ref{eq:B}}),
between the normalized magnetization $m$ and the normalized applied
field $h$, shown vs.\ the applied field frequency $\nu$, for
$T$$=$$20\,{\rm K}$ and $100\,{\rm K}$. The bars are {\em not} error
bars; instead, they represent the {\em standard deviation} of $B$ at
$T$$=$$100\,{\rm K}$. The standard error in the measured mean is
smaller than the symbol size. As $B$ is the reactive part of the
nonlinear system response, the first zero crossing is taken as a sign
of resonance. A particular example of $m(t)$ (solid curve) and
$h(t)/10$ (dotted curve) for $T$$=$$100\,{\rm K}$ and $\nu=0.2\,{\rm
GHz}$ is shown for several hysteresis-loop periods in the inset.}
\end{figure}

\newpage

\begin{figure}[tb]
\center{\epsfxsize 4 in \epsfbox{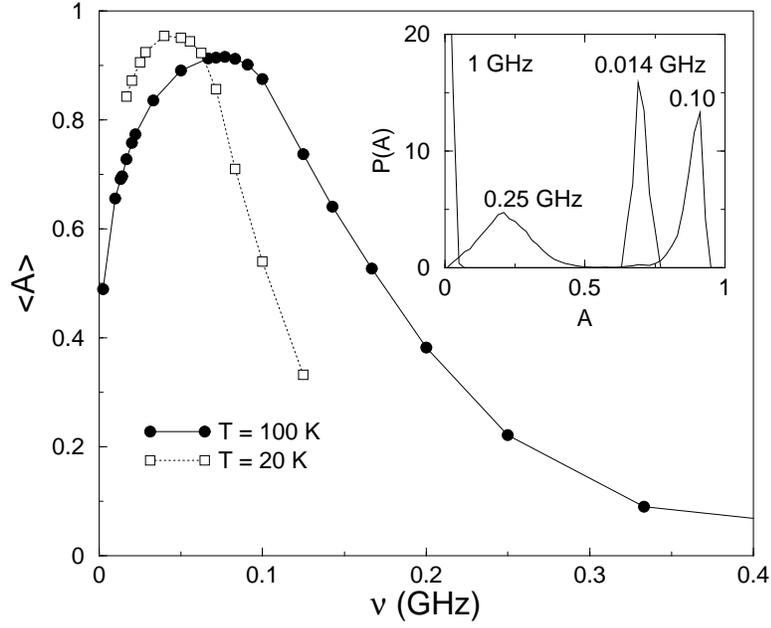}}
\caption[]{ The mean hysteresis-loop area $A$, defined in
Eq.~(\ref{eq:A}), vs.\ the applied field frequency $\nu$ for
$T$$=$$20\,{\rm K}$ and $100\,{\rm K}$. The standard error in the
measured mean is smaller than the symbol size.  The maximum in $A$,
which is the dissipative part of the nonlinear response function, is
associated with resonance. The temperature-dependent resonance
frequencies agree well with those indicated in Fig.~1. Histograms of
$A$ for several different applied field frequencies at $100\,{\rm K}$
are shown in the inset.}
\end{figure}

\newpage

\begin{figure}[tb]
\center{\epsfxsize 4 in \epsfbox{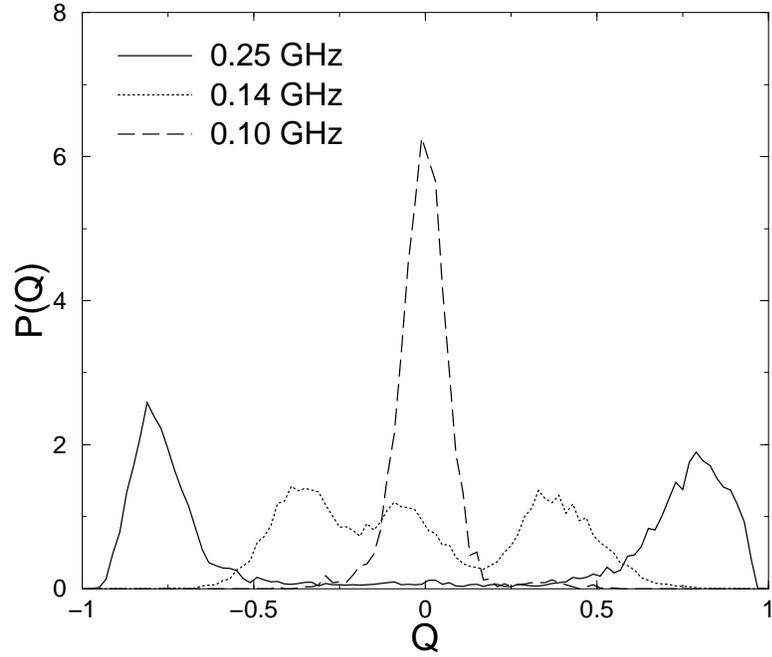}}
\caption[]{ Histogram of the period-averaged magnetization $Q,$
defined in Eq.~(\protect{\ref{eq:Q}}), for applied field frequencies
of $0.25$, $0.14$, and $0.1\,{\rm GHz}$. $Q$ is a dynamic order
parameter for the system. The two peaks a the highest frequency
indicate a broken symmetry state.  }
\end{figure}

\end{document}